\documentclass[
 rsi,
amsmath, amssymb,
 reprint,
]{revtex4-1}

\usepackage{graphicx}
\usepackage{dcolumn}
\usepackage{bm}
\usepackage[colorlinks=true, citecolor=red, linkcolor=black, urlcolor=black]{hyperref}

\usepackage[utf8]{inputenc}
\usepackage[T1]{fontenc}
\usepackage{mathptmx}
\usepackage{etoolbox}

\makeatletter
\def\@email#1#2{%
 \endgroup
 \patchcmd{\titleblock@produce}
  {\frontmatter@RRAPformat}
  {\frontmatter@RRAPformat{\produce@RRAP{*#1\href{mailto:#2}{#2}}}\frontmatter@RRAPformat}
  {}{}
}%
\makeatother
\begin{document}


\title{
Preparing State-Selected Molecular Beams with high Efficiency using Quantum Cascade Lasers
}

\author{Ardra P. S.}

\author{Pranav R. Shirhatti\textsuperscript{*}}
\email{pranavrs@tifrh.res.in}

\affiliation{Tata Institute of Fundamental Research Hyderabad,\\
36/P Gopanpally, Hyderabad 500046, Telangana, India}


\begin{abstract}
We demonstrate the ro-vibrational state-selective preparation of CO molecules ($v$ = 0, J = 0 $\rightarrow$ $v$ = 1, J = 1) in an internally cold molecular beam using a continuous-wave, narrow-linewidth distributed feedback type quantum cascade laser, with an output wavelength of 4.65 $\mu$m.
Using a modulation-free wavelength stabilization scheme, we show that population transfer to near saturation levels can be obtained with sufficient long term stability, as needed for state-resolved scattering experiments.  
We show preliminary results of CO molecules prepared in the $v$ = 1 state, scattering from a clean Cu(100) surface.
This study showcases the promise of QCL based excitation for state-selected molecular beam based scattering experiments, especially involving transitions in the mid-infrared domain ($\lambda$ > 4.5 $\mu$m) which are otherwise difficult to access using conventional sources.\end{abstract}

\maketitle

\section*{\label{sec:level1}INTRODUCTION}

Molecular beam based scattering measurements have played a foundational role in advancing our understanding of intermolecular interactions, collisional energy transfer, and dynamics of chemical reactions.
In particular, quantum state resolved scattering measurements, provide information on the nature and magnitude of collisional energy transfer, and on the role of internal excitation on reaction barrier crossing rates of elementary chemical reactions.
These measurements typically employ internally cold molecular beams with optical state preparation and detection using laser spectroscopy based methods.
Ro-vibrational state-selective preparation of incident molecules has been of particular interest for molecule-surface scattering studies. 
These studies have revealed the nature of vibrational energy dissipation pathways, especially looking at the coupling between nuclear and electronic degrees of freedom \cite{rahinov2024,Alec_review_2015}, and mode selective nature of chemical reactions on surfaces \cite{juurlink_uTZ_rEV2009, Beck_Chadwick_rev_2016}.

Ro-vibrational state-specific preparation of internally cold molecules in a molecular beam requires intense, narrow linewidth and stable radiation sources which can be used for direct infrared (IR) absorption or excitation using stimulated Raman pumping (SRP).
Several different approaches have been developed in this regard.
Multimode pulsed IR generated using difference frequency generation has been used for measuring NO ($v$ = 1) relaxation upon scattering from a LiF(100) surface \cite{NO_lif,Misewich1985}.
Similar methods have also been used to study the vibrational to translational energy transfer in collisions of vibrationally excited HCl from Au(111) surface \cite{rahinov_2008_HCL_Au}. 
Beck and co-workers have studied mode specific dissociation of vibrationally excited CH$_4$, prepared using an intense, multimode excitation source, on a Ni(100) surface \cite{ex3MethaneNi,RB_2003_RSI}.
Excitation schemes using single longitudinal mode IR radiation with active stabilization, which are much more efficient compared to multimode sources, have also been developed and used in several state-selected scattering studies \cite{reisler_2000_hcl_mgo,NO_vib_tr1,NO_vib_tr2}.
The high intensity, narrow linewidth and single longitudinal mode nature of the output obtained using specially designed optical parametric oscillator/amplifier (OPO/A) systems are of particular interest for such measurements, as it provides high excitation efficiency, even for weak overtone transitions and offers a wide output wavelength range \cite{dean_guyer1993,OPO_SLM_2017,Avinash}.
 SRP method has been used for looking at state-selective scattering of H$_2$ from Cu surfaces \cite{sitz2001_h2_cu_2_1,sitz_h2cu_raman,sitz_Rev_2002}.
These methods have also been developed further to go past the saturation limit \cite{sarp_zare,XUEMING_SARP}. 
Recently, very high intensity OPO/A based sources for Raman pumping of extremely weak overtone transitions have also been demonstrated \cite{xueming_CJCP, Xueming_RSI2020}.

While the pulsed laser based excitation methods are very versatile in terms of the accessible wavelength range, their major drawback is that only a small fraction of the overall molecular beam is excited. 
Excitation using continuous-wave (cw) sources (as opposed to pulsed) overcomes this limitation.
Gough and co-workers \cite{miller_1978_NO} have demonstrated ro-vibrational excitation of NO in a molecular beam using a cryogenically cooled lead salt diode laser along with bolometric detection.
McCabe and co-workers \cite{Utz_RSI_2000}, using color center lasers have shown that up to 15 - 25\% of the total CH$_4$ molecules can be prepared in the vibrationally excited $\nu_3$ state. 
This was used to study the mode selective dissociation of CH$_4$ on Ni(100) surface.
Similar studies were also performed by Higgins and co-workers \cite{scoles_cw} using an optical build up cavity for exciting the weak 2$\nu_3$ mode for state-selective dissociation of CH$_4$ on Pt(111) surface.
Subsequent experiments in this regard have benefited immensely from the development of high power (Watt-level) cw OPOs with output in the 2 - 4.3 $\mu$m wavelength range \cite{Beck_UTZ_2013_review,Beck_Chadwick_rev_2016}.
The combination of high power and narrow linewidth (1 MHz) has also been used to perform complete population transfer using adiabatic passage techniques \cite{beck_RAP,juurlink_co2}.
Recently, using a combination of narrow linewidth near-IR diode lasers and very high finesse optical build up cavities, ro-vibration state preparation of extremely weak transitions such as CO $v$ = 0 - 3 and HD $v$ = 0 - 2 \cite{xueming} in a molecular beam has also been demonstrated. 

While these developments using cw excitation methods have proved to be extremely useful, the low peak power (compared to pulsed sources) limits output wavelength range to regions only where suitable low loss gain media are available.
This poses a challenge for extending state-selective preparation methods to a wider range of molecules, especially where the transitions lie in the mid-IR region, beyond 4.5 $\mu$m. 
Quantum cascade lasers (QCLs) are an attractive alternative for exciting molecular fundamental vibrational transitions in the mid-IR. 
Their compact footprint, typically a few square centimeters, output wavelengths in the mid-IR region, near room temperature operation  and output powers ranging from a few mW to hundreds of mW make them attractive for integration with molecular-beam experiments.
Despite these advantages, applications of compact QCL sources for state-selective molecular-beam experiments remain relatively few.
Recently, Krohn and Chandler \cite{olivia1,olivia} used QCL for state-selective excitation combined with ultra high-resolution velocity selection, producing NO molecules with a velocity spread of $\sigma$ < 3 m/s.  
Insero and co-workers \cite{borri} demonstrated high-resolution ro-vibrational spectroscopy of metastable CO in a molecular beam using a frequency-stabilized QCL with 11 digits of precision. 
Timothy and co-workers \cite{david_nesbit_qcl} demonstrated direct absorption measurements of molecules in a molecular beam using a QCL in a multipass arrangement, achieving an absorbance sensitivity of $<10^{-5}$. 

Given the promising nature of QCLs, in this work we demonstrate the preparation of a ro-vibrationally excited $^{12}$CO molecular beam by exciting the fundamental ro-vibrational transition R(0), at 2147.08 cm$^{-1}$ ($v$ = 0, J = 0 $\rightarrow$ $v$ = 1, J = 1) using a compact, low-power QCL with active wavelength stabilization. 
We provide a detailed account of the excitation scheme and wavelength stabilization method used. 
We have characterized the state-selectively prepared molecular beam to quantify the population transfer fraction  and evaluate the short and long-term stability of our excitation scheme.
Our measurements showcase that this compact QCL based excitation source for state-selected preparation of CO can achieve population transfer close to the saturation limit. 
Finally, we show some preliminary results of CO ($v$ = 1) scattering from a clean Cu(100) surface, demonstrating the utility of this method for state-selected molecular beam-surface scattering experiments.

\subsection*{\label{sec:level2}METHODS}
The mid-IR radiation source employed in these experiments was a distributed-feedback (DFB) type QCL (Alpes Lasers, Model No. CW-SM-Q-2147.08-20-0.3-HHL-L).
This QCL has a center wavelength corresponding to 2147 cm$^{-1}$, close to the $v$ = 0, J = 0 $\rightarrow$ $v$ = 1, J = 1 transition at 2147.08 cm$^{-1}$. 
The laser emits 36 mW of power (measured using a pyroelectric sensor S401C, Thorlabs) at the center wavelength when operated with an injection current of 173.3 mA and maintained at a temperature of 300 K (TC-3, temperature controller, Alpes laser). 
It has a mode hop free scanning range of > 4 GHz.
The output beam is slightly elliptical, with diameters (full width half maximum, FWHM) of 2.6 mm (horizontal) and 3.0 mm (vertical) at a distance of 102 mm from the QCL and half-angle divergences of 0.32 mrad and 0.63 mrad, respectively.

\begin{figure*}\includegraphics[width=\textwidth]{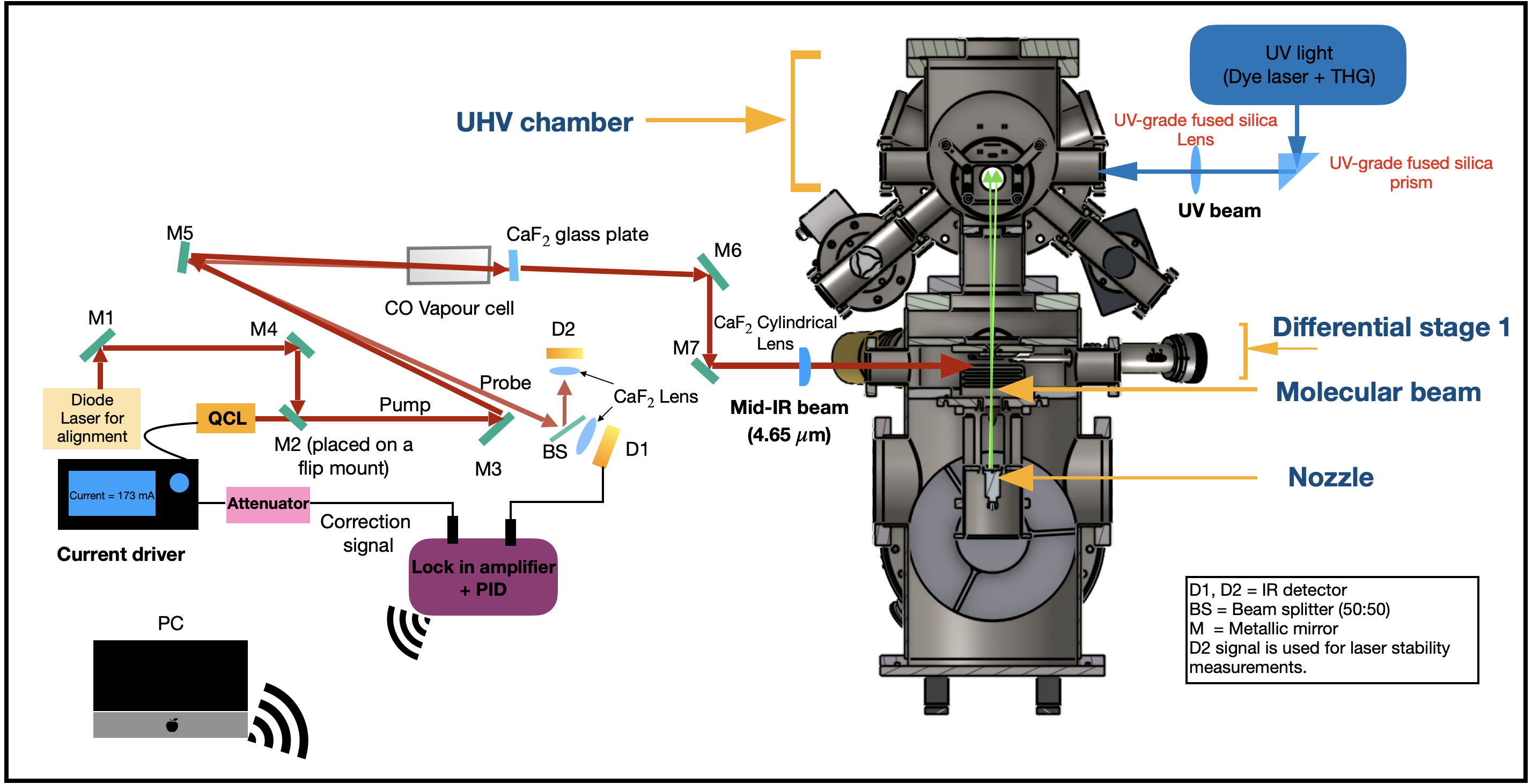}
\caption{\label{Schematic}Schematic diagram of the experimental setup showing the SAS arrangement used for laser frequency locking and the molecular beam experimental setup.}
\end{figure*}
The current and temperature tuning coefficients of the QCL were determined to be 1.05 GHz/mA and 4.53 GHz/K. 
This was done by measuring the absorption features corresponding to the $^{13}$CO R(14) transition (2147.20 cm$^{-1}$) and the $^{12}$CO R(0) transition (2147.08 cm$^{-1}$) using current and temperature tuning (SI-1). 
The transition frequencies were obtained from the HITRAN database \cite{hitran}.
Since our objective was to achieve an efficient and steady population transfer in the molecular beam, a modulation-free locking scheme for the QCL wavelength stabilization was chosen. 
This was implemented by locking the QCL output wavelength to the side of the Doppler-free saturation absorption peak of the $v$ = 0 - 1, R(0) transition, in a low pressure gas cell (20-40 $\mu$bar, path length 320 mm). 
Under these conditions, about 30\% of the IR intensity was absorbed by the CO gas in the cell and remaining 70\% was utilized for the molecular beam excitation.
The pump and probe beams used for the saturation absorption spectroscopy (SAS) were aligned in a near-collinear counter propagating configuration to minimize back reflections into the QCL (Fig. \ref{Schematic}). 
This was essential as optical isolators are not readily available in this wavelength range.
The probe beam was a reflection from an uncoated CaF$_2$ plate ($\sim$10\%).
The resulting SAS signal was detected using an IR detector (PDA07P2, Thorlabs, 9 MHz bandwidth).
The overall optical alignment was done by co-aligning a red diode laser along the QCL beam path.

The error signal corresponding to the frequency drift was obtained using a proportional-integral-derivative (PID) controller of the Laser lock box (LLB, Moku:Go, Liquid instruments). 
The correction signal was applied  to the external modulation input of the current driver (QCL500 LAB, Wavelength Electronics) through an impedance-matched switchable attenuator (SI-2).
The effective voltage-to-frequency tuning coefficients after attenuation were 54.8 MHz/V and 95.1 MHz/V. The larger coefficient was used to scan the entire Doppler-broadened spectrum and the smaller coefficient was used for frequency correction. 
The external voltage-to-frequency coefficient was determined by comparing voltage and temperature scan based measurement of the Doppler-broadened spectrum and using the known temperature-to-frequency coefficient.
Further details are provided in SI-3.
To minimize electrical noise and avoid ground loops, the LLB was powered using a battery, and communication with the PC was established via WiFi instead of USB or Ethernet connections. 

Once the wavelength was stabilized, the QCL beam was directed into the first differential pumping stage of the molecular beam-surface scattering setup, where it intersected the molecular beam (Fig. \ref{Schematic}).
The $v$ = 1 signal was optimized by slightly tilting the QCL beam to compensate for the offset between the locking point and the transition peak.
A detailed description of this molecular beam-surface scattering setup can be found elsewhere \cite{saurabh,ThesisSKS,ThesisAKS}.
Briefly, a CO molecular beam was generated by a pulsed solenoid valve (Part No. 009-1643-900, Parker), driven using an Iota One valve driver (Parker Hannifin).
The valve was operated at a repetition rate of 10 Hz with an electrical pulse duration of 250 $\mu$s. 
It has an orifice diameter of 1 mm and was operated at a stagnation pressure of 4 bar and a temperature of 300 K.
To minimize back reflections into the QCL, the optical windows of the differential stage were tilted by 3$^\circ$ downward with respect to the lab vertical plane.
A cylindrical lens was used to focus the IR beam along the vertical direction (lab frame) to maximize its overlap with the molecular beam.
At the intersection region with the molecular beam, the laser beam had horizontal and vertical spatial dimensions (FWHM) of 5.3 mm and 1.0 mm, respectively (SI-4).

The excited CO molecules were detected 300 mm downstream (UHV chamber) using a 2+1 REMPI scheme \cite{rempi_ground} via the   X$^1\Sigma^+$ ($v$ = 1, J = 1) $\rightarrow$ B$^1\Sigma^+$ ($v$ = 1, J = 1) transition.
The UV radiation for REMPI was generated using a pulsed dye laser (Lioptec, LiopStar-E $\&$ SHG, THG), pumped by a Nd$^{3+}$:YAG laser (Continuum Surelite III laser), operating at a repetition rate of 10 Hz with a pulse duration of 5 ns. 
The dye laser fundamental output at 690 nm (Pyridine 1 in methanol) was frequency  converted to its third harmonic at 230 nm through second-harmonic generation and sum-frequency generation ($\omega + 2\omega$) using beta barium borate crystals.
The UV beam (0.5 to 1 mJ/pulse) was focused onto the molecular beam using a 500 mm focal length lens. 
The resonantly ionized molecules were extracted using a repeller plate and accelerated through a  field free time-of-flight tube.
\begin{figure}
    \centering
    \includegraphics[width=\linewidth]{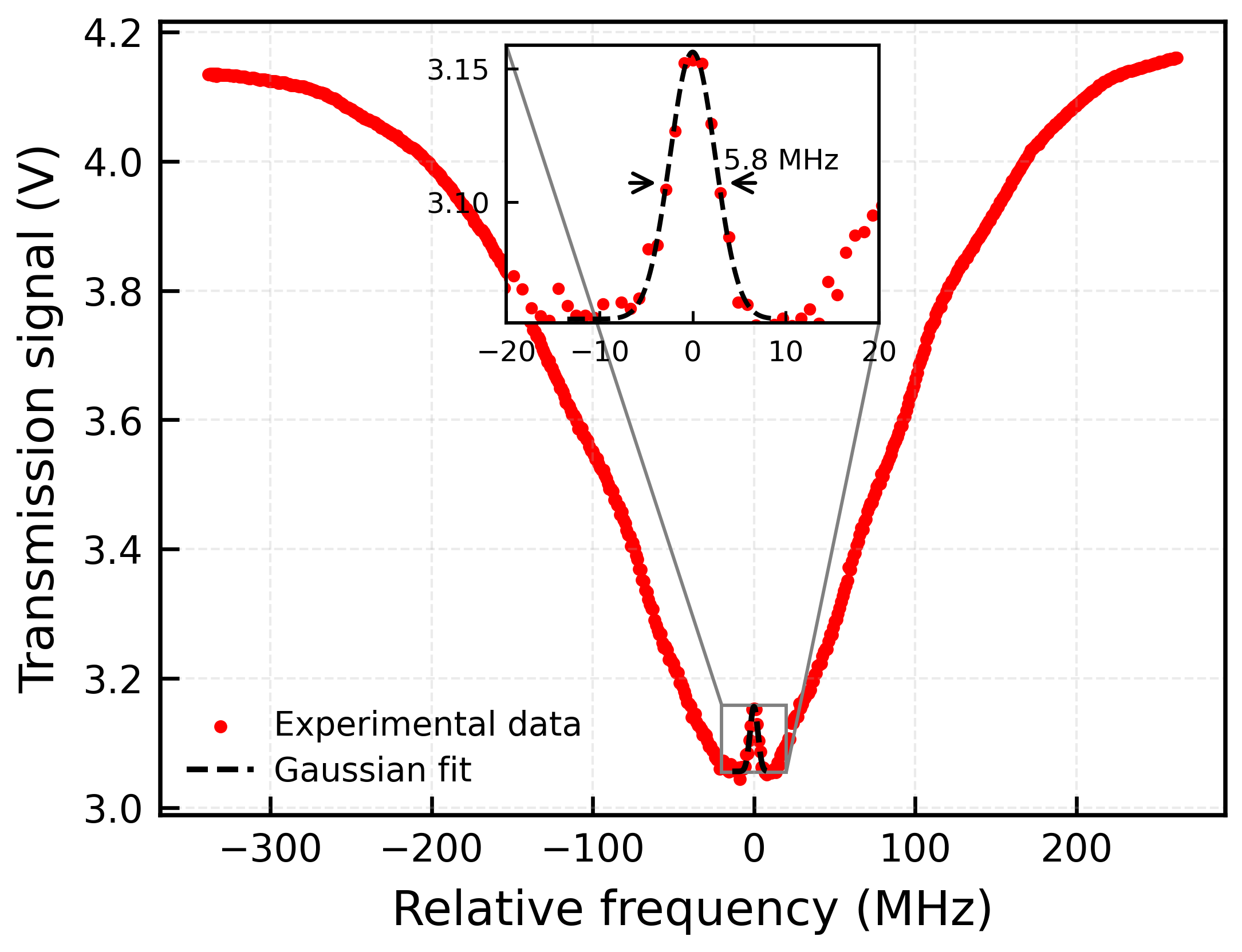}
    \caption{Doppler-broadened $^{12}$CO R(0) transition along with the Doppler-free spectrum obtained using SAS method (red points). 
    The inset provides an enlarged view of the Doppler-free peak and the corresponding fit (black dashed line), showing a FWHM of 5.8 MHz.
    }
    \label{lamb dip}
\end{figure}
The ions were detected using a dual-stack microchannel plate detector (40mm diameter, F2225-21P, Hamamatsu). 
The resulting signal was imaged on a phosphor screen (ProxiVision P43) and recorded using a global-shutter camera  (Arducam 2 MP OV2311 monochrome UVC) synchronized with the laser pulses.
For scattering studies of state-selectively prepared CO molecules on Cu(100) surface, a single-crystal Cu(100) (MaTecK Material-Technologie $\&$ Kristalle GmbH), with dimensions of 10 mm $\times$ 10 mm $\times$ 2 mm, and cut to an accuracy of $0.1^\circ$ was used.
The surface was prepared by Ar$^{+}$ sputtering for 40 min at an ion energy of 3 keV and an ion current of 0.8 $\mu$A.
Subsequently, the surface cleanliness was checked using Auger electron spectroscopy (AES). 
Once cleaned, the surface was annealed at approximately 830 K for about 30 min, followed by cooling to 300 K.
The scattering measurements were done under ultra high vacuum conditions, with a chamber base pressure of 1.5 $\times$ 10$^{-9}$ mbar, under which the surface remained clean, with no detectable contamination on the surface for > 5 hours (as confirmed by AES).
The CO beam was made incident along the surface normal of the clean Cu(100) surface. 
The scattered and incident molecules in the $v$ = 0 and $v$ = 1 states were detected by measuring the corresponding REMPI signals.

\section*{Results and Discussions}\label{sec:results}
\begin{figure}
    \centering
    \includegraphics[width=\linewidth]{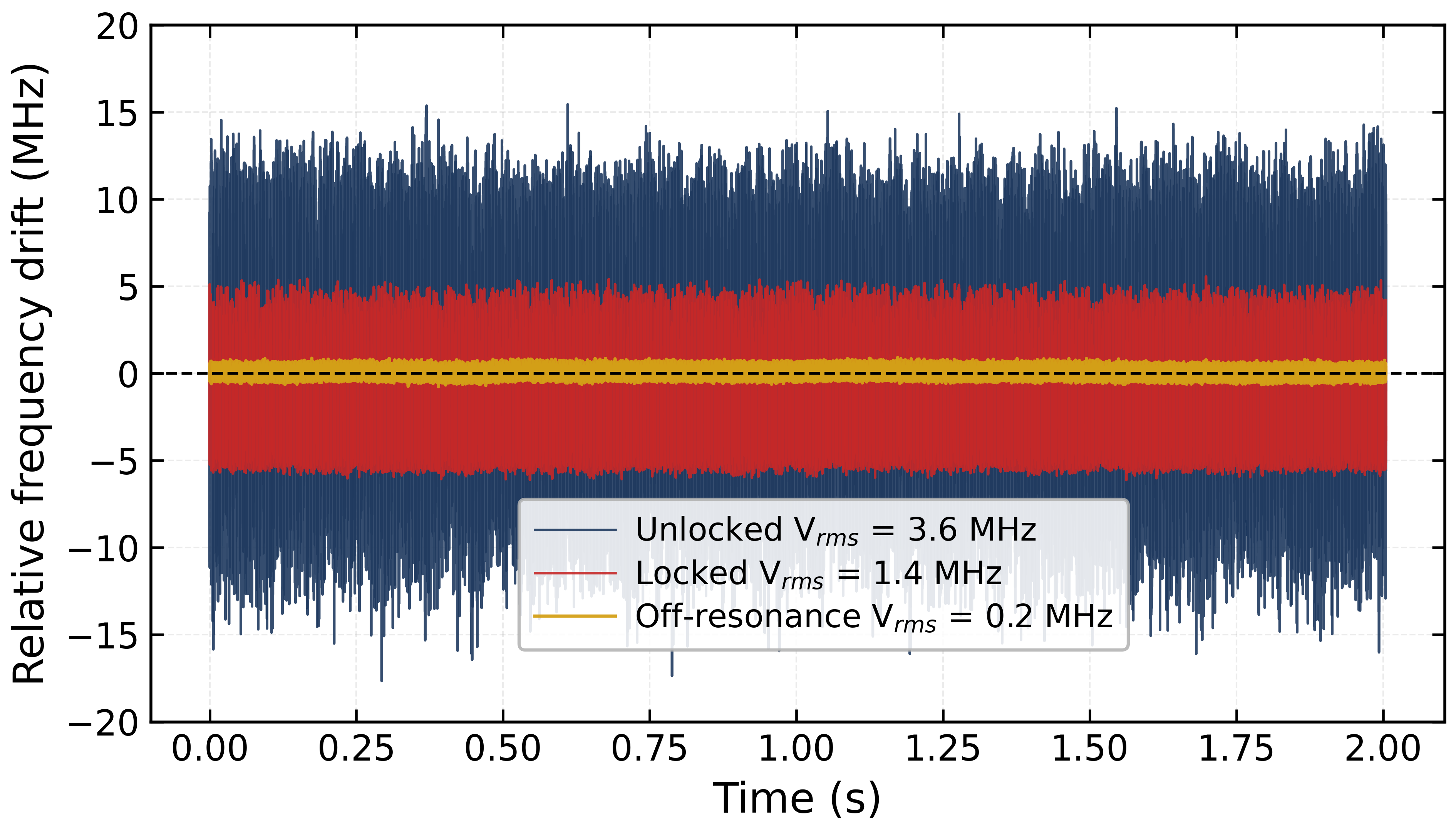}
    \caption{
    The blue curve represents the relative frequency drift with time, obtained under free running condition of the QCL. 
    The RMS fluctuation was 3.6 MHz in this case.
    The red curve shows the same under locked conditions and has an RMS fluctuation of 1.4 MHz.
    The yellow curve represents the system noise measured at off resonance conditions.
    This corresponds to an equivalent RMS frequency fluctuation of 0.2 MHz.
    }
    \label{stability}
\end{figure}

DFB QCLs are highly sensitive to thermal and electronic disturbances, and to achieve long term stability, these need to be minimized. 
As mentioned earlier, absorption spectrum of CO molecules in a cold molecular beam have a narrow line width of the order of few MHz 
resulting from the beam divergence (residual Doppler broadening), which makes it necessary to have good short and long term stability of the order of 1 MHz.
Figure \ref{lamb dip} shows the Doppler-broadened and the Doppler-free spectra. 
The Doppler-free spectra has a linewidth of 5.8 MHz (FWHM) and a contrast of 10$\%$ relative to the Doppler-broadened profile.
This was used as the frequency reference for locking the laser. 
The observed linewidth of the Doppler-free profile is largely dominated by the frequency fluctuations in free running condition of the QCL and some power broadening (SI-5).
For frequency stability measurements, the probe beam was divided into two equal parts using a 50:50 beam splitter (BSW511, Thorlabs), and each beam was detected separately using IR detector (PDA07P2, Thorlabs). 
The output signals from D1 and D2 (Fig. \ref{Schematic}) were used for laser locking and monitoring the frequency drifts, respectively. For stability measurements, the signal from D2 was amplified using a low-noise preamplifier (SR560, Stanford Research Systems) with a gain of 10$\times$ and recorded using an oscilloscope (Moku:Go, Liquid Instruments).
The short-term laser frequency stability was measured under both locked and unlocked conditions at a sampling rate of 1.7 MSps.
The free-running stability was measured with the QCL wavelength set to the side of the Doppler-broadened spectrum, while the stability under locked condition was measured on the side of the Doppler-free peak. 
The voltage-to-frequency calibration for each of these discriminator curves was taken into account while evaluating the stability (SI-6).
The results of stability measurements are shown in Fig. \ref{stability}. 
The root mean square (RMS) frequency fluctuation was measured to be 1.4 MHz under locked conditions, compared with 3.6 MHz in the unlocked condition. 
The intrinsic noise of the measurement system was equivalent to 0.2 MHz (estimated from off resonance measurements). 

The rotational temperature of our molecular beam was determined to be 8 - 12 K, corresponding to 30 - 21\% of the CO molecules populating the X$^1\Sigma^+$ ($v$ = 0, J = 0) state. 
For a two-level system interacting with radiation, with spontaneous emission rate being negligibly slow, the excited state population fraction is given by \cite{Utz_RSI_2000}
\begin{equation}
    f_2 = \frac{g_2}{g_1+g_2}\left\{1-exp[-\rho B_{21}(\frac{g_1+g_2}{g_1})t]\right\}
    \label{eq:popfraction}
\end{equation}
Here, B$_{21}$ is the Einstein B coefficient, $\rho$ is the radiation density, and t is the time of interaction of the molecules with the radiation. 
As the interaction time, radiation density, or both increase, the exponential term approaches zero, and the population fraction reaches its saturation value
${g_2}/({g_1+g_2})$.
For the case of R(0) transition used here, this corresponds 75$\%$, which means that a maximum overall population transfer of 16 - 22\% under our measurement conditions.

\begin{figure}
    \centering
    \includegraphics[width=\linewidth]{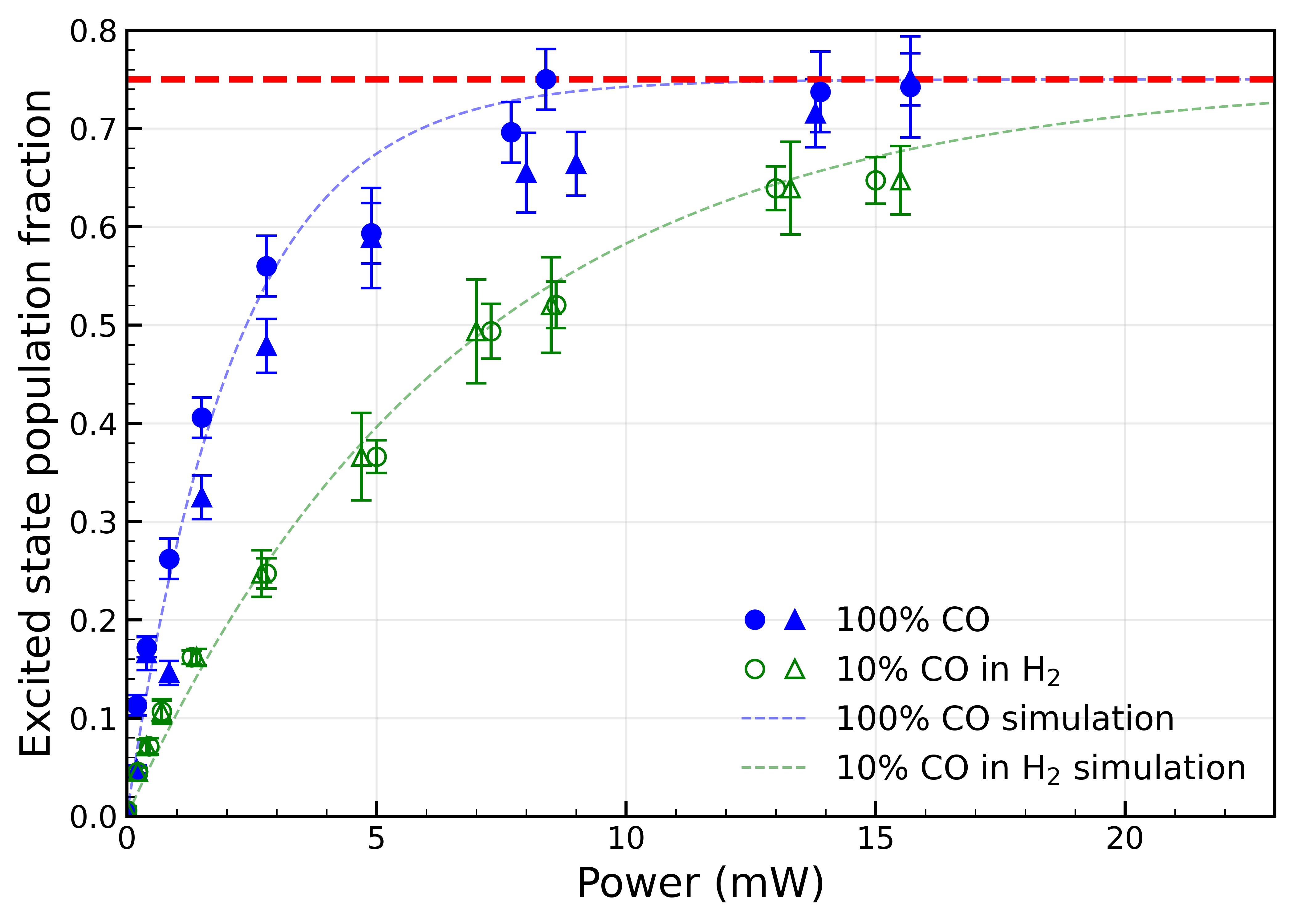}
    \caption{
    REMPI signal for the $v$ = 1 transition as a function of IR excitation power from the QCL. 
    Blue points correspond to two independent measurements using a beam of 100$\%$ CO.
    Green points correspond to two independent measurements using a beam of 10\% CO in H$_2$.
    The curves represent the estimates using equation [\ref{eq:popfraction}].
    The red dashed line denotes the saturation value of 0.75 for R(0) transition considered here.
    }
    \label{saturation_curve}
\end{figure}

In order to quantify the population transfer observed in our experiments, we measured the IR power dependence of the $v$ = 1 REMPI signal (Fig. \ref{saturation_curve}).
For lower powers, the $v$ = 1 signal showed a linear dependence and eventually resulting in saturation at larger powers, as expected from equation \ref{eq:popfraction}. 
Measurements were performed using two different molecular beams: a 100$\%$ CO beam with a velocity of 557 $\pm$ 21 m/s having an interaction time of 9.5 $\mu$s, and a 10$\%$ CO in H$_2$ beam with a velocity of 1894 $\pm$ 28 m/s having an interaction time of 3 $\mu$s. 
As shown in Fig. \ref{saturation_curve}, the $v$ = 1 signal saturates beyond $\sim$ 10 mW for the pure CO beam. 
In the case of faster beam, excitation fraction of 65$\%$ (87$\%$ of the saturation level) was achieved. 
These results clearly demonstrate that efficient population transfer is achieved using this set up.

\begin{figure}
    \centering
    \includegraphics[width=\linewidth]{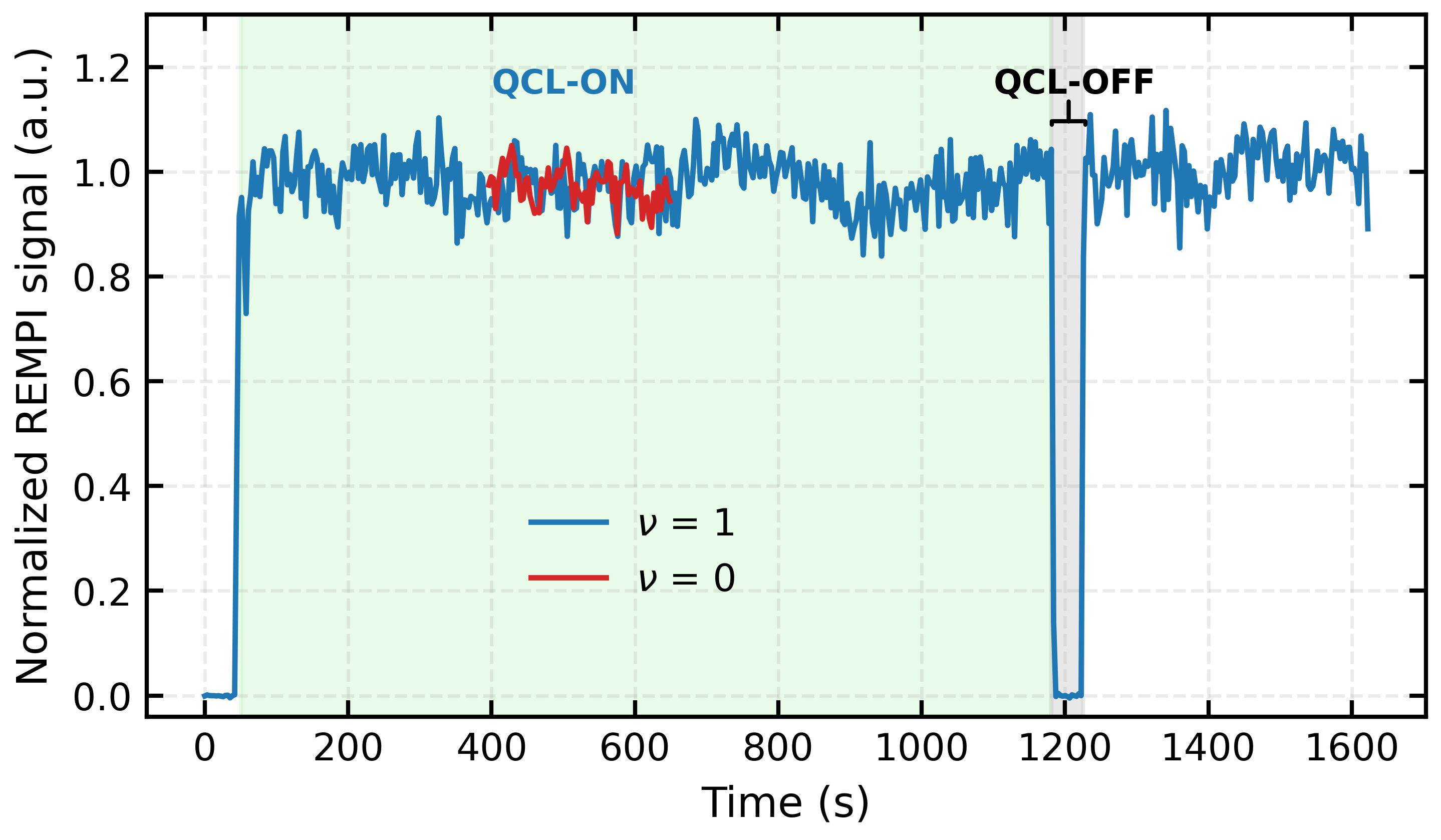}
    \caption{
    REMPI signal for CO $v$ = 1 (blue) and 0 (red) measured as a function of time to evaluate the long term performance of the QCL based excitation scheme.
    The relative fluctuations of the $v$ = 1 and 0 signals were observed to be comparable, having values of 3.8\% and 3.2\%, respectively.
    Around the 1200 sec mark, the QCL was blocked and the signal drops to zero. This shows that the entire $v$ = 1 signal observed is from the population in the $v$ = 1 state created by the IR excitation.
    }
    \label{rempi}
\end{figure}

Another essential feature needed for using this excitation scheme in scattering experiments is long term stability, typically over a time scale exceeding 1000 sec.
This was measured by monitoring the $v$ = 1 signal for over 1500 sec using REMPI 
(Fig. \ref{rempi}). 
In addition to the wavelength instability of the IR beam, shot-to-shot variations in gas density and laser pulse energy also contribute to the overall fluctuations observed in the $v$ = 1 signal (blue curve, Fig. \ref{rempi}).
To quantify these, REMPI signal vs. time for the $v = 0$ (red curve, IR off, Fig. \ref{rempi}) and $v$ = 1 (blue curve, Fig.  \ref{rempi}) were compared. 
The fluctuations in the $v$ = 0 signal provides a measure of the intrinsic stability of our overall measurement system.
The RMS fluctuations for the $v$ = 0 and $v$ = 1 were observed to be 3.2$\%$ and 3.8 $\%$ respectively. 
Given that these are comparable, we conclude that the fluctuations in $v$ = 1 signal arise primarily from the detection method rather than the wavelength instability of  the IR source.

\begin{figure*}[t]
     \includegraphics[width=\linewidth]{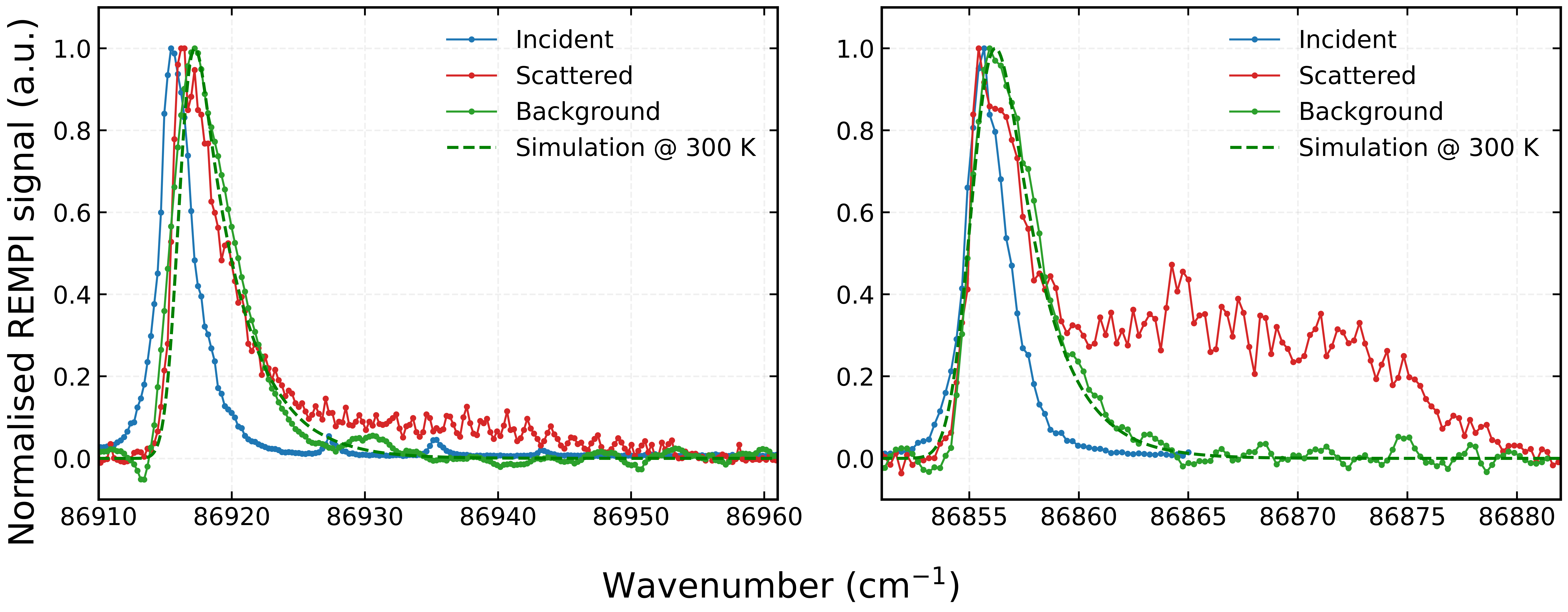}
    \caption{Left: Peak normalized REMPI spectra of incident, scattered, and background CO molecules in the $v$ = 0 state. 
    Right: Peak normalized REMPI spectra of incident, scattered, and background CO molecules in the $v$ = 1 state.
    The measured background spectra look similar to the simulated spectra with a rotational temperature of 300 K (green dashed lines). 
    The enhanced population of higher rotational states in the scattered spectra for both $v$ = 0 - 0 (left) and $v$ = 1 - 1 (right) channels indicate hyperthermal rotational energy distribution, characteristic of direct inelastic scattering.}
    \label{scatter}
\end{figure*}

Spatial profiles of the $v$ = 0 and $v$ = 1 signals show a good overlap both along the laser propagation axis and the axis along the vertical plane of the laboratory frame (SI-7).
This indicates that the entire molecular beam is getting excited uniformly.
The temporal profile of our molecular beam shows a peak followed by a tail at longer delays (SI-7). 
We note that the $v$ = 1 excitation efficiency in this tail region is lower than that compared to the peak.
This can potentially arise from the inhomogeneities in the velocity (Doppler shifts) and rotational state distribution (population in J = 0) across the temporal profile of the molecular beam pulse.
Our analysis of velocity and rotation state distributions suggests that this effect arises largely from the later (SI-7).
It should be noted that these effects reflect the limitation of present molecular-beam source and not the intrinsic limitation of QCL based excitation technique.
This non-uniformity can be addressed by using a continuous nozzle. 
For experiments which require a pulsed molecular beam, a molecular-beam chopper \cite{chopper} can be used to select a narrower and more homogeneous part of the pulse. 
Alternatively, a high-performance nozzle \cite{LV_valve} can provide improved rotational cooling and a uniform rotational population distribution. 
Improved rotational cooling would also increase the population in the J = 0 states, thereby increasing the fraction of molecules available for excitation and improving the overall excitation efficiency.
For example, at rotational temperatures of $\sim$ 5 K  \cite{xueming}, the population in the $J = 0$ state can reach $\sim$ 46\%. Thus, at saturation, the overall fraction of the population in the $v$ = 1 state would be 35\%.

We now present some preliminary results of CO, $v$ = 0 - 0 and $v$ = 1 - 1 scattering from a clean Cu(100) surface.
Here, a 10$\%$ CO in H$_2$ molecular beam with an energy of 0.52 eV was sent at normal incidence on a clean Cu(100) surface. 
Figure \ref{scatter} shows the normalized REMPI spectra of the incident and scattered CO molecules, together with the background for both $v$ = 0 - 0 (left) and $v$ = 1 - 1 (right) channels.
The incident molecules (blue curve) show a narrow rotation state distribution as expected. 
The scattered molecules (red curve) show a much broader distribution extending to higher J states, indicating hyperthermal rotational excitation, characteristic of direct inelastic scattering.
In addition, we observed a thermal-like distribution (green curve). 
We attribute this to the thermalized background signal from molecules undergoing multiple collisions with the grids and chamber walls.
This background rotational distribution is consistent with the simulated population distribution at 300 K (PGOPHER \cite{pgopher}, molecular constants obtained from the NIST Chemistry WebBook).
A quantitative determination of the vibrational survival probability will require systematic measurements of the incident and scattered ($v$ = 1) populations at different collision energies, accounting for background contributions, REMPI detection efficiencies, spatial variations in the MCP gain, and differences in the incident and scattered molecular-beam distributions. This will be pursued in independent future work.

\section*{Summary and Concluding Remarks}\label{sec:summary}
The QCL-based excitation system described in this work enables efficient ro-vibrational state-selective preparation of molecules in  a molecular beam.
The preparation scheme, together with its characterization, population-transfer measurements, and frequency stabilization, have been demonstrated in detail. 
The use of a cw QCL enables continuous excitation of fundamental vibrational transitions, and is especially beneficial in the cases where transitions lie in the mid-IR wavelength region. 
The modulation-free, side-of-fringe frequency-locking scheme used here provides frequency stabilization without introducing additional modulation-related frequency fluctuations.
Using this approach we demonstrated saturation of the transition in the molecular beam. 
These results clearly demonstrate that the QCL-based state preparation can be integrated with molecular-beam scattering measurements. 
Further improvements to this experimental system could include locking the laser to a stable reference cavity instead of an absorption line. 
This will allow for frequency scanning while maintaining the laser lock and only requires < 1 mW power, compared to the $\sim$10 mW required to saturate the signal for the SAS locking scheme.
In addition, the use of external-cavity QCLs \cite{david_nesbit_qcl} in place of DFB QCLs will provide a wider tuning range ($\sim$ 100 cm$^{-1}$), allowing excitation of additional rotational transitions.
Independently, in future work, we aim to quantify the survival probability of the vibrationally excited molecules as a function of collision energy, providing a more detailed understanding of vibrational energy transfer.

\section*{Supplementary information}\label{sec:SI}
\begin{itemize}
    \item SI-1: Determining the current-to-frequency and temperature-to-frequency tuning coefficients 
    \item SI-2: Attenuator circuit 
    \item SI-3: Determining the external voltage to frequency tuning coefficients 
    \item SI-4: IR beam shape in the region of intersection with the molecular beam
    \item SI-5: Power broadening
    \item SI-6: Calibration for the instantaneous stability measurements
    \item SI-7: Spatial and temporal profiles of molecules in the $v$ = 0 and $v$ = 1 states
\end{itemize}

\section*{Acknowledgments}\label{sec:akhnowlegment}
We acknowledge the funding provided by the Department of Atomic Energy, Government of India, under Project Identification No. RTI 4007, 
Anusandhan National Research Foundation (ANRF), Science and Engineering Research Board (SERB) and Department of Science and Technology, India (Grant number. CRG/2022/002943 and ANRF/ARG/2025/005588/CS). 
We thank Olivia Krohn (Sandia National Laboratories,  USA) and Athanasios Laliotis (Université Sorbonne Paris Nord) for insightful discussions on electronic noise reduction in the QCL control system. 
We also thank G. Rajalakshmi (TIFR Hyderabad) for valuable discussions on laser locking and Vishal Ranjan (TIFR Hyderabad) for providing the low noise preamplifier. 
We acknowledge Minali (TIFR Hyderabad) for her assistance in the REMPI measurements and Sai Vijay Mocherla  (TIFR Hyderabad) for help in configuring the LLB with battery operation.
Finally, we acknowledge the members of the institute mechanical workshop for fabricating several custom-designed components used in our experimental setup.
\vspace{1em}
\begin{center}
    \textbf{AUTHOR DECLARATIONS}
\end{center}

\noindent\textbf{Conflict of Interest}

The authors have no conflicts to disclose.

\vspace{0.5em}

\noindent\textbf{Author Contributions}

APS and PRS designed the overall experimental strategy.
APS performed the estimations,  designed the necessary components, characterized the performance the QCL system, performed the measurements and analyzed the data with inputs from PRS.
PRS conceptualized and supervised the project. 
APS and PRS discussed the results, analysis, and contributed to writing the manuscript.

\bibliography{aipsamp}

\end{document}


\newpage
\maketitle
\tableofcontents
\thispagestyle{fancy}
\fancyhf{}
\renewcommand{\footrulewidth}{0.4pt}
\fancyfoot[L]{%
$\ast$ Author to whom correspondence should be addressed.\\
e-mail: pranavrs@tifrh.res.in
}

\newpage

\section{SI-1: Determining the current-to-frequency and temperature-to-frequency tuning coefficients }

\begin{figure}[h]
    \centering

    \begin{minipage}{0.495\linewidth}
        \centering
        \includegraphics[width=1\linewidth]{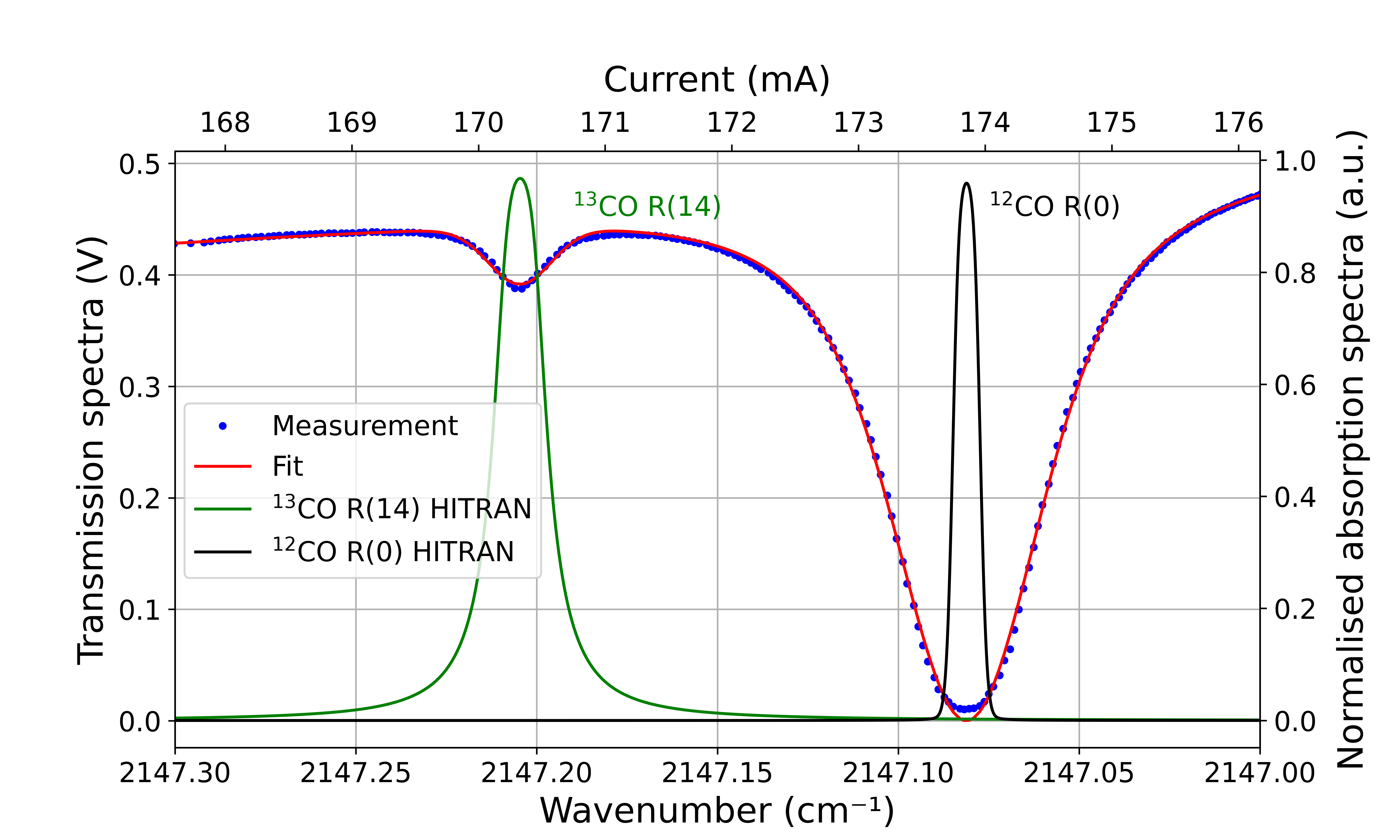}
    \end{minipage}%
    \hspace{0.005\linewidth}%
    \begin{minipage}{0.495\linewidth}
        \centering
        \includegraphics[width=1\linewidth]{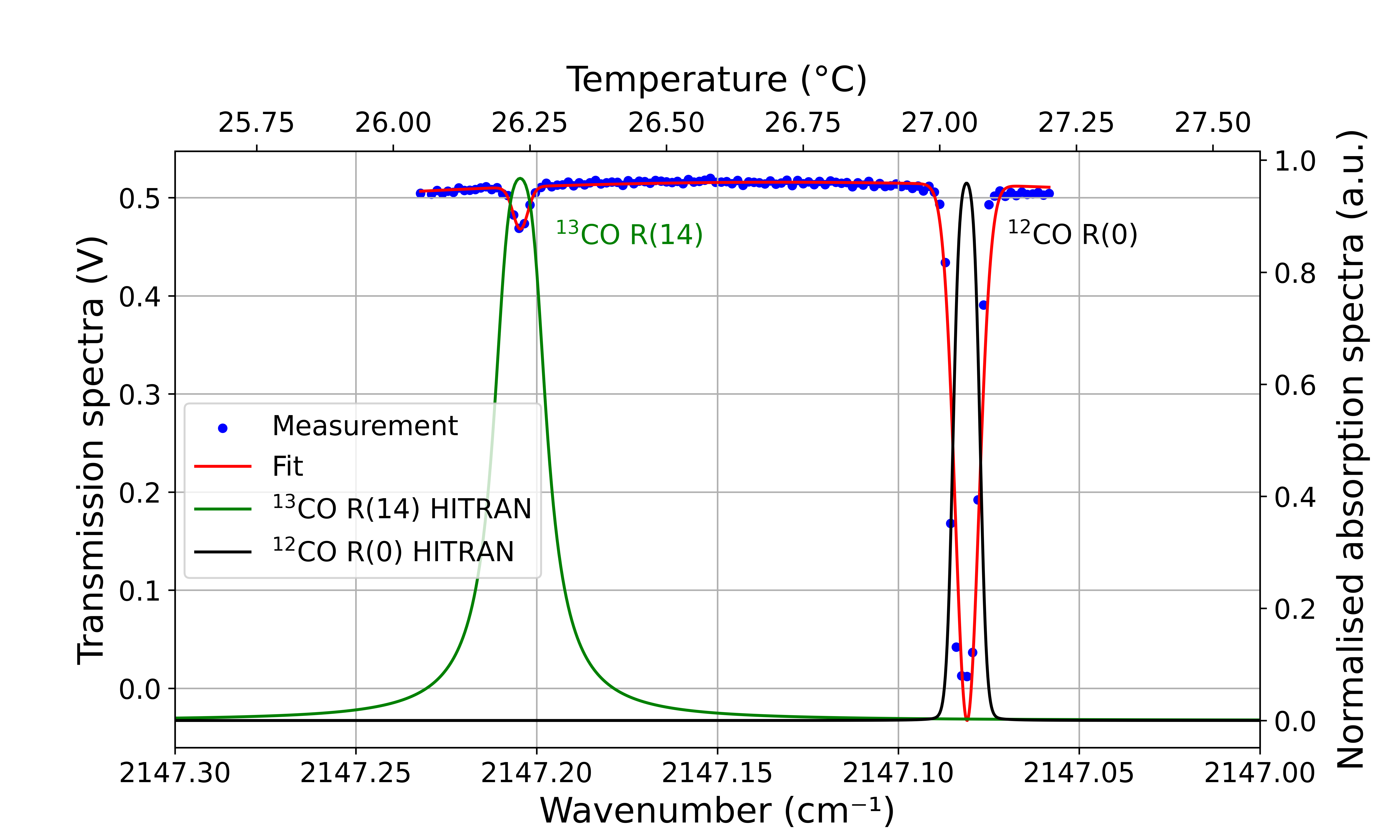}
    \end{minipage}
    \caption{Left: Transmission spectrum of CO in a gas cell (300 K, 10 mbar), obtained by scanning the injection current of the QCL, while keeping the temperature constant at 300 K. Right: Transmission spectrum of CO in a gas cell (300 K, 0.46 mbar) obtained by scanning the temperature of the QCL. The injection current was maintained constant at 173.3 mA.
    The two peaks (red dotted curve) observed correspond to $^{13}$CO R(14) and $^{12}$CO R(0) transition lines and the corresponding HITRAN \cite{hitran1,hitran2} simulations are shown as green and black curves. 
    }
    \label{fig:i_T_to_nu}
\end{figure}
The temperature-to-frequency $ \left(d\nu/dT \right) $ and current-to-frequency $ \left(d\nu/dI \right) $ tuning coefficients were determined by analyzing the $^{13}$CO R(14) and $^{12}$CO R(0) (Fig. \ref{fig:i_T_to_nu}) transition lines.
The peak-to-peak separation in frequency, between these transitions is 3.7 GHz, as obtained from HITRAN database \cite{hitran1,hitran2}. 
The change in injection current between the same two peaks was measured to be 3.53 mA, yielding a current-to-frequency tuning coefficient of (3.7 GHz)/(3.530 mA) = 1.05 GHz/mA. 
Similarly, the difference in the temperature was measured to be 0.82 K, corresponding to a temperature-to-frequency tuning coefficient of (3.7 GHz)/(0.82 K) = 4.53 GHz/K. 
The current and temperature were scanned by the external control of the current driver and the temperature controller using a LabVIEW (2021, National Instruments) program. 

\section{SI-2: Attenuator circuit}
\begin{figure}[H]
    \centering
    \includegraphics[width=0.9\linewidth]{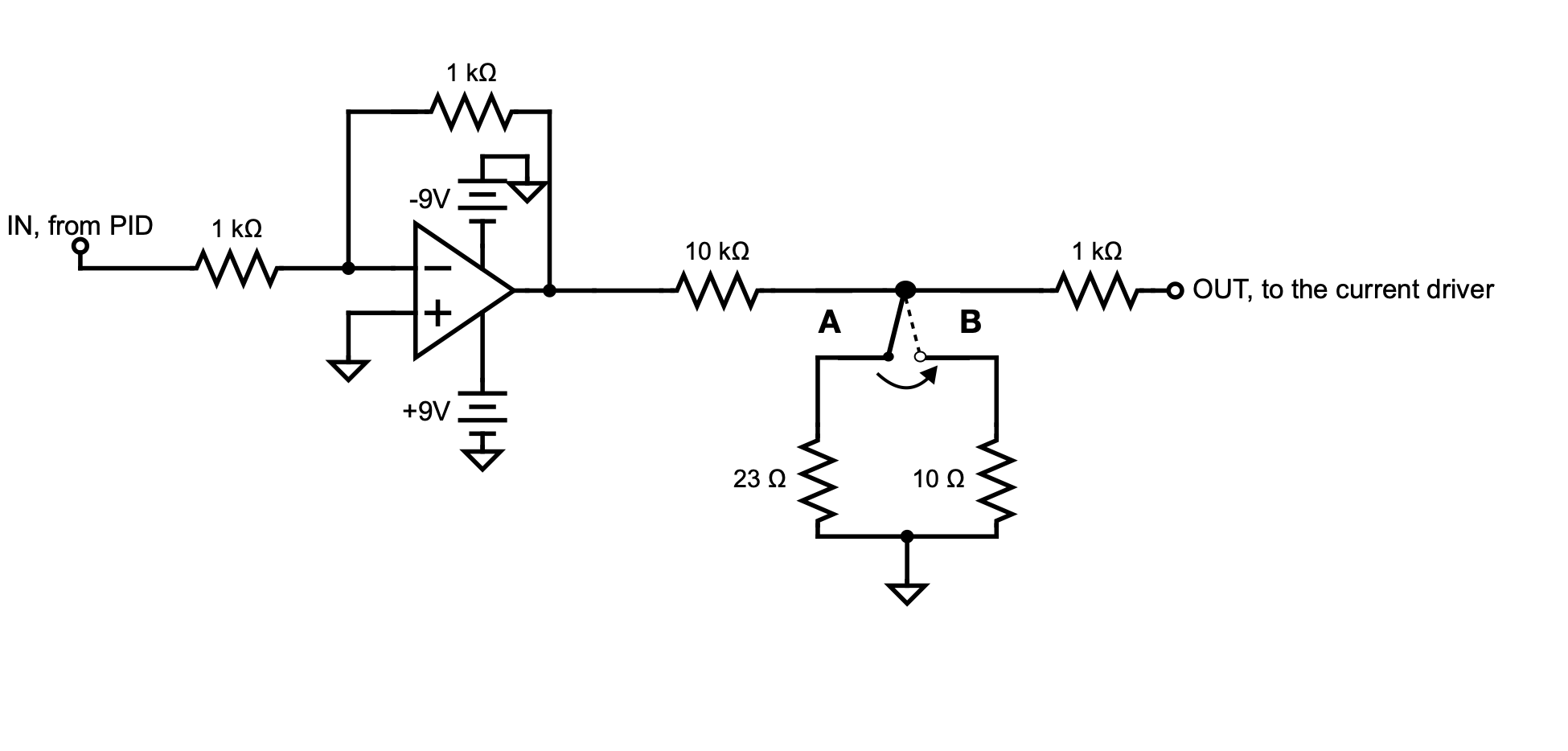}
    \caption{Circuit schematic of the attenuator used for injecting the correction signal from LLB into the QCL current driver.}
    \label{fig:circuit}
\end{figure}

The external control input of the current driver has a transfer coefficient of \(0.1~\mathrm{A/V}\). 
Combined with the current-to-frequency tuning coefficient of \(1.05~\mathrm{GHz/mA}\), this means that a 1 MHz frequency change requires a current change of approximately \(1~\mathrm{\mu A}\), corresponding to a voltage change of \(10~\mathrm{\mu V}\) at the external control input. 
Therefore, MHz level frequency corrections require voltage control at the microvolt level. 
To provide small voltage changes reliably, a larger input voltage was attenuated and used for external control with the desired resolution.
The attenuator circuit consists of an input buffer that receives signal from the Laser Lock Box (LLB, Moku:Go, Liquid Instruments).
The output from the buffer is then sent to a voltage divider for attenuation. 
A single-pole double-throw switch was used to select one of the two attenuator branches: Branch A for large frequency steps and branch B for small frequency steps (Fig. \ref{fig:circuit}). 
The output impedance of the attenuator was set to 1 k$\Omega$ to match the input impedance of the current driver’s external modulation input.

\section{SI-3: Determining the external voltage to frequency tuning coefficients}
Frequency scan and the feedback for correcting frequency drifts are implemented by varying the QCL current through the external control input of the current driver.
This external voltage is attenuated before being applied to the external control input.
Therefore, the voltage-to-frequency conversion factor must be determined to quantify the corresponding frequency changes.
The procedure for determining the voltage-to-frequency tuning coefficients for both attenuator settings (Fig. \ref{fig:circuit}) is summarized in the flowchart below.
\begin{figure}[htbp]
    \centering
    \begin{tikzpicture}[
        node distance=7mm,
        box/.style={
            rectangle,
            rounded corners=2pt,
            draw=black,
            align=center,
            text width=15cm,
            minimum height=1.5cm,
            inner sep=6pt,
            font=\small
        },
        arrow/.style={
            -{Stealth[length=2mm]},
            thick
        }
    ]

    \node[box] (step1) {
        \textbf{Temperature-to-frequency tuning coefficient}\\
        $\frac{d\nu}{dT}$ = 4.53 GHz/K (SI-1)
    };

    \node[box, below=of step1] (step2) {
        Doppler-broadened spectra obtained from the QCL temperature scan and external-voltage (with attenuation, large-step setting) scan were overlaid (Fig. \ref{fig:calibratin_attenuator}, left). From the overlaid spectra, the frequency shift produced by a temperature change of 0.021 K was matched to that produced by a 1 V change in the external voltage.\\
        $\displaystyle \frac{dT}{dV}$ = 0.021 K/V
    };

    \node[box, below=of step2] (step3) {
        \textbf{Calculate the voltage-to-frequency coefficient}\\
        $\displaystyle
        \left(\frac{d\nu}{dV}\right)_{\mathrm{large-step}}
        =\frac{d\nu}{dT}\frac{dT}{dV}
        =95.1\ \mathrm{MHz/V}$
    };

    \node[box, below=of step3] (step4) {
        \textbf{Transfer calibration from large to small-step setting of the attenuator}\\
        The Doppler-free spectra obtained using both attenuator step settings (Fig. \ref{fig:calibratin_attenuator}, center and right) were compared to transfer the calibration from the large-step to the small-step setting. Since both spectra were recorded sequentially under the same experimental conditions, the Doppler-free linewidth was assumed to remain unchanged, yielding
        $\displaystyle
       \left(\frac{d\nu}{dV}\right)_{\mathrm{small-step}} = 54.8\ \mathrm{MHz/V}$\\
       
    };

    \draw[arrow] (step1) -- (step2);
    \draw[arrow] (step2) -- (step3);
    \draw[arrow] (step3) -- (step4);

    \end{tikzpicture}
    
    \label{fig:voltage_calibration_flowchart}
\end{figure}
\newpage
\begin{figure}[H]
    \centering
    \includegraphics[width=1\linewidth]{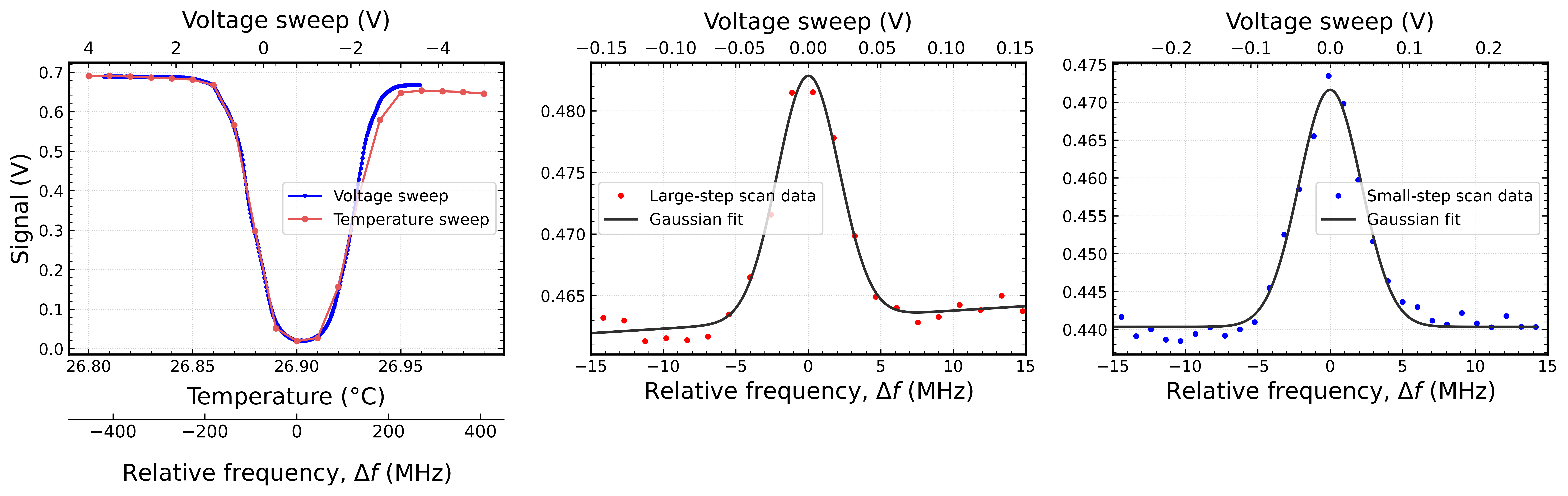}
    \caption{
    Left: The red points show the $^{12}$CO R(0) transmission spectrum obtained from a temperature sweep, while the blue points shows the corresponding spectrum obtained from a voltage sweep with the attenuator set to the large-step setting (Fig. \ref{fig:circuit}). During these measurements, the gas cell pressure was $1.2\times10^{-1}$ mbar. The two spectra were overlaid to establish the voltage to temperature and subsequently voltage to frequency conversion factor, using d$\nu$/dT = 4.53 GHz/K (Fig. \ref{fig:i_T_to_nu}). 
    Center and right: Doppler-free spectra obtained using larger and smaller voltage step scans of the attenuator (Fig. \ref{fig:circuit}), respectively. The black curves represent fits corresponding to the Doppler-free spectra.}
    \label{fig:calibratin_attenuator}
\end{figure}
\section{SI-4: IR beam shape in the region of intersection with the molecular beam}
\begin{figure}[H]
    \centering
    \includegraphics[width=0.8\linewidth]{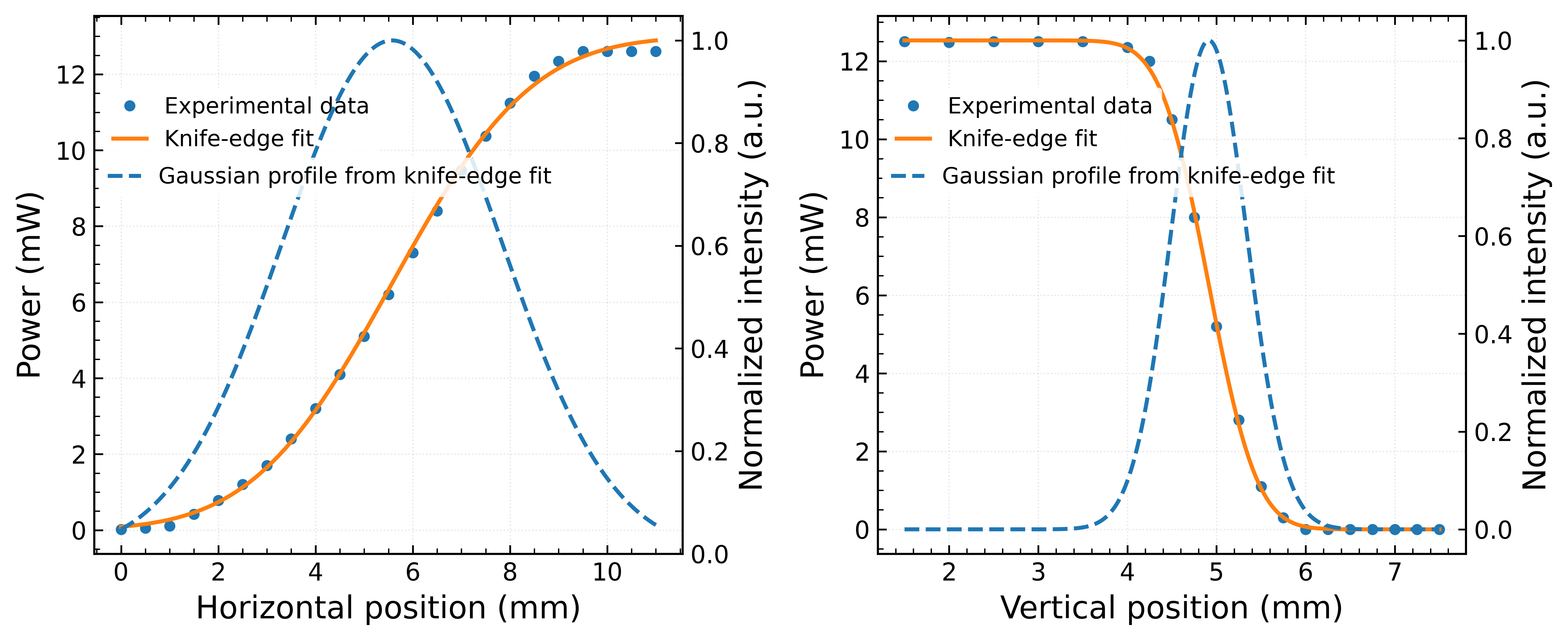}
    \caption{Results of knife edge scan measurements for both horizontal (left) and vertical plane (right).
    Blue dots represent the knife-edge measurement data, and blue dashed lines show the corresponding Gaussian fit. Left: profile along the molecular beam propagation direction at the point of intersection. Right: profile along the vertical direction in the laboratory frame at the same location.}
    \label{fig:knife}
\end{figure}
\begin{figure}[H]
    \centering
    \includegraphics[width=0.8\linewidth]{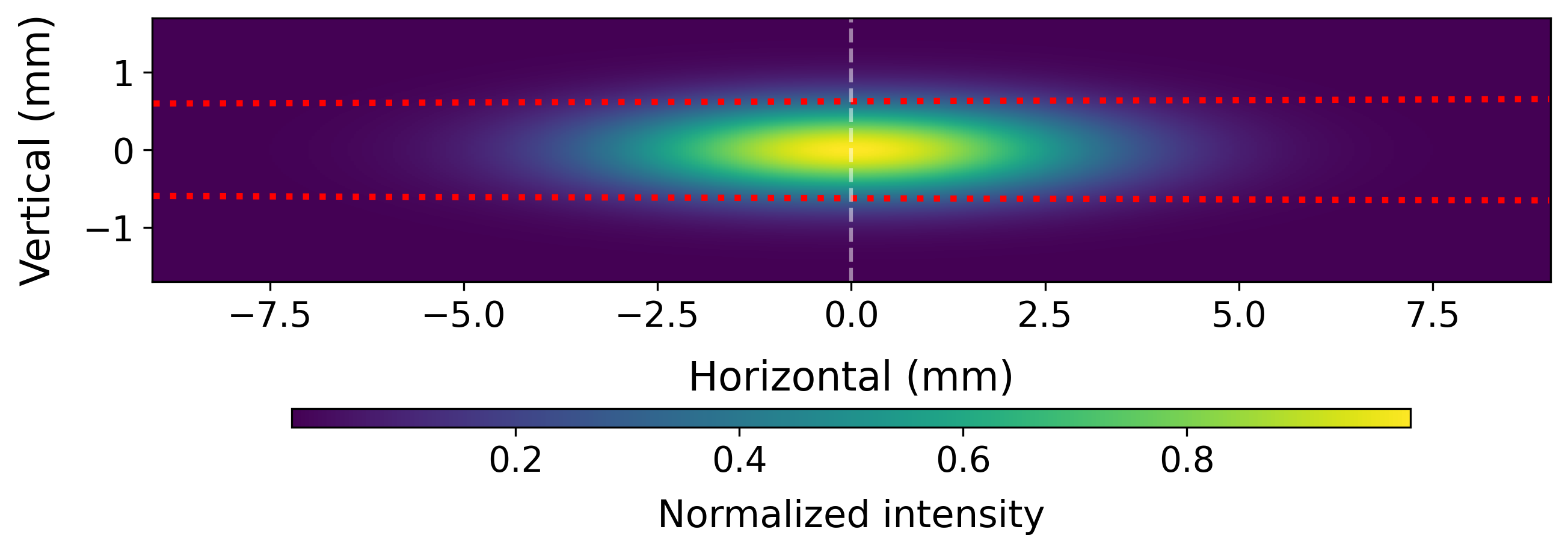}
    \caption{A plot showing the IR laser beam profile in the region where it intersects the molecular beam. The red dotted line corresponds to the subsequent apertures down stream and provide an estimate of the molecular beam size}
    \label{fig:beam dia}
\end{figure}

In order to achieve a good overlap among the IR and the molecular beam the laser beam was focused using a cylindrical lens before intersecting the molecular beam, resulting in an elliptical beam profile with focusing along the vertical plane in the lab frame (Fig. \ref{fig:beam dia}). 
The beam diameter at the point of intersection was measured using a knife-edge scan (Fig. \ref{fig:knife}). 
The measured full width half maximum of the beam profile were 5.3 mm and 1.0 mm along the horizontal and vertical planes of lab frame, respectively. The overall IR beam profile is shown in Fig. \ref{fig:beam dia}

\section{SI-5: Power broadening}
\begin{figure}[h]
    \centering
    \includegraphics[width=1\linewidth]{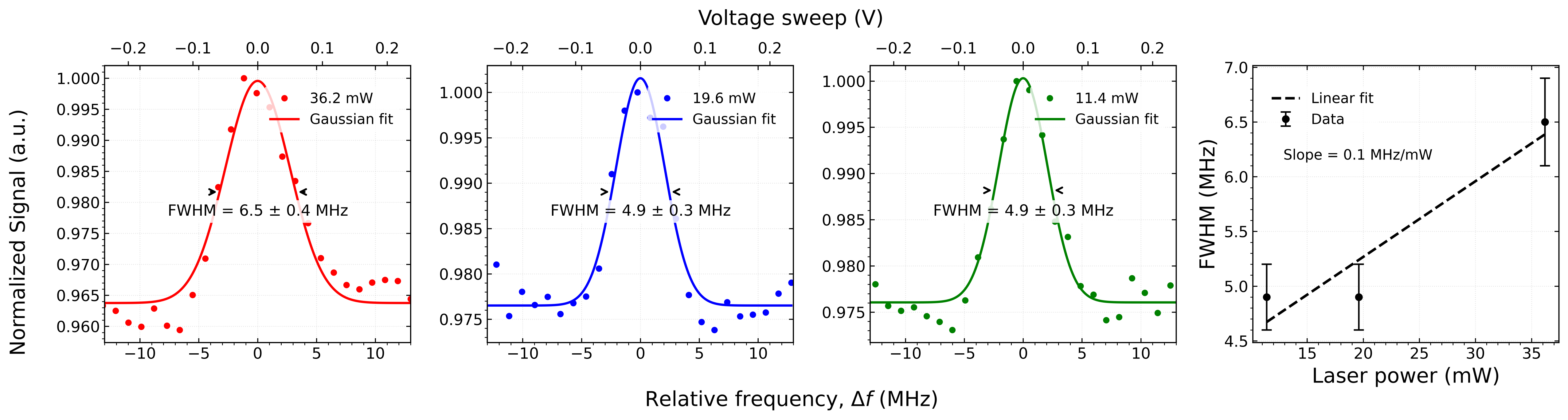}
    \caption{Measurement of the Doppler-free peak at different pump powers: 36.2 mW (left, red curve), 19.6 mW (middle left, blue curve), and 11.4 mW (middle right, green curve). The corresponding linewidths are 6.9 MHz, 4.9 MHz, and 4.9 MHz, respectively. Right: Linewidth of the Doppler-free spectrum as a function of laser power.}
    \label{fig:power_b}
\end{figure}

To determine power broadening, the Doppler-free peak was measured at different IR pump powers. 
The power entering the vapor cell was varied by placing cover slips acting as attenuators (CS, germanium-borosilicate glass, thickness: 0.13 mm to 0.17 mm) in the beam path.
At the maximum pump power of 36.2 mW, measured without CS, the Doppler-free linewidth was 6.9 MHz (Fig. \ref{fig:power_b}, red curve, left).
Inserting a CS into the beam path reduced the pump power to 19.6 mW, yielding a linewidth of 4.9 MHz (Fig. \ref{fig:power_b}, blue curve, middle left). A further reduction in pump power to 11.4 mW did not lead to any additional narrowing, with the linewidth remaining at 4.9 MHz (Fig. \ref{fig:power_b}, green curve, middle right).
The initial linewidth reduction indicates a contribution from power broadening. Below 19.6 mW, the linewidth remains unchanged, suggesting that power broadening is negligible and the linewidth is primarily limited by laser frequency fluctuations.
The linewidth measured at maximum power was slightly larger than the Doppler-free linewidth reported in the main manuscript (Fig. 2, main manuscript).
This difference is due to an additional source of mechanical noise from the vacuum pump connected to the gas cell.
For the measurements reported in the main manuscript, the vacuum pump was turned off while making the measurements.
From the measured linewidth at different IR pump power, we estimate a power-broadening coefficient of 0.1 MHz/mW (Fig. \ref{fig:power_b}, right).
\clearpage
\section{SI-6: Calibration for the instantaneous stability measurements}
\begin{figure}[h]
    \centering
    \includegraphics[width=0.5\linewidth]{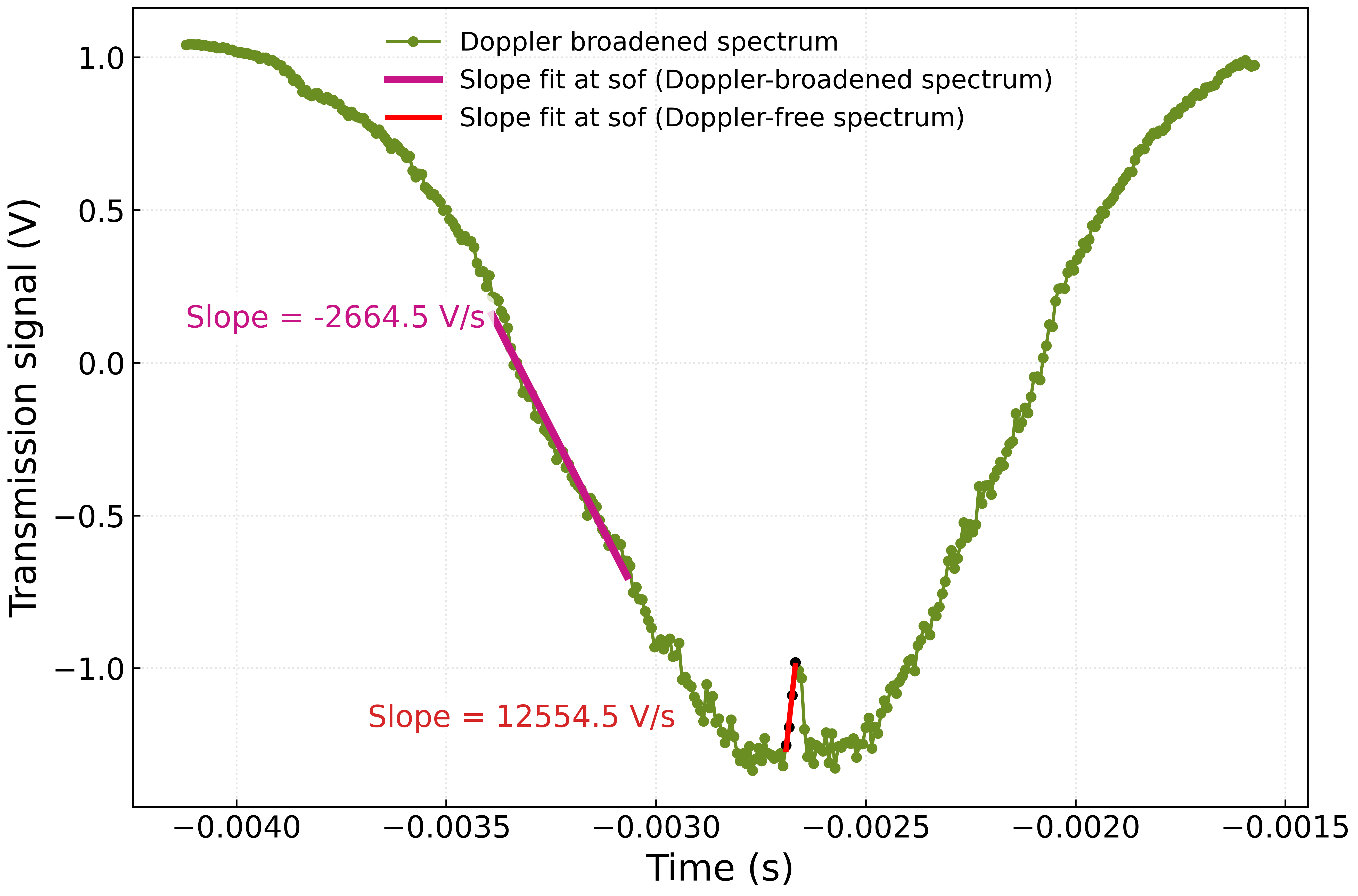}
    \caption{The green spectrum shows the Doppler-broadened R(0) line of $^{12}$CO. The purple line represents the slope at the side of fringe (sof) used to determine the free running laser linewidth. The red line represents the slope at the sof of the Doppler-free peak, where the stability under locked condition is measured.}
    \label{fig:calibration_stability}
\end{figure}

The flowchart below summarizes the procedure for determining the voltage-to-frequency conversion coefficients at the of side-of-fringe (sof) slopes of the Doppler-broadened and Doppler-free peaks.
These slopes were used as frequency discriminators to measure the instantaneous stability of the QCL under free-running and locked conditions, respectively.

\begin{figure}[htbp]
    \centering
    \begin{tikzpicture}[
        node distance=7mm and 5mm,
        box/.style={
            rectangle,
            rounded corners=2pt,
            draw=black,
            align=center,
            text width=7.0cm,
            minimum height=1cm,
            inner sep=5pt,
            font=\small
        },
        arrow/.style={
            -{Stealth[length=2mm]},
            thick
        }
    ]

    \node[box, text width=9cm] (start) {
        \textbf{Voltage sweep for obtaining the R(0) line}\\
        Sweep rate = $3600\ \mathrm{V/s}$\\
        Voltage-to-frequency coefficient =
        $54.8\ \mathrm{MHz/V}$ (SI-2)
    };

    \node[box, below left=of start, xshift=3.0cm] (doppler) {
        Slope measured at the sof of the 
        Doppler-broadened line\\ =
        $2664.5$ V/s (Fig. \ref{fig:calibration_stability}, violet curve)\\
    };

    \node[box, below right=of start, xshift=-3.0cm] (lamb) {
        Slope measured at the sof of the 
        Doppler-free line
        = $12554.5$ V/s  (Fig. \ref{fig:calibration_stability}, red curve)
    };

    \node[box, below=of doppler] (free) {
        \textbf{Discriminator slope in terms of frequency}\\
        
        $\displaystyle
        \frac{3600}{2664.5}\times54.8
        \approx74\ \mathrm{MHz/V}$
    };

    \node[box, below=of lamb] (locked) {
        \textbf{Discriminator slope in terms of frequency}\\
        $\displaystyle
        \frac{3600}{12554.5}\times54.8
        \approx16\ \mathrm{MHz/V}$
    };

    \draw[arrow] (start.south) -- (doppler.north);
    \draw[arrow] (start.south) -- (lamb.north);
    \draw[arrow] (doppler) -- (free);
    \draw[arrow] (lamb) -- (locked);

    \end{tikzpicture}
    
    \label{fig:frequency_correction_calibration}
\end{figure}

\clearpage
\section{SI-7: Spatial and temporal profiles of molecules in the
$v$ = 0 and $v$ = 1 states}
\subsection{Spatial profile}
\begin{figure}[htbp]
    \centering
    \begin{minipage}[c]{0.3\linewidth}
        \centering
        \includegraphics[width=\linewidth]{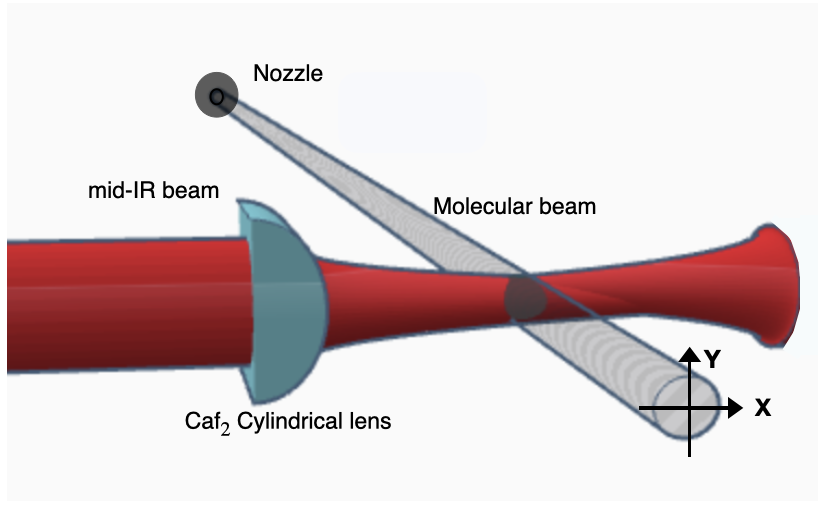}
    \end{minipage}
    \hspace{0.01\linewidth}
    \begin{minipage}[c]{0.55\linewidth}
        \centering
        \includegraphics[width=\linewidth]{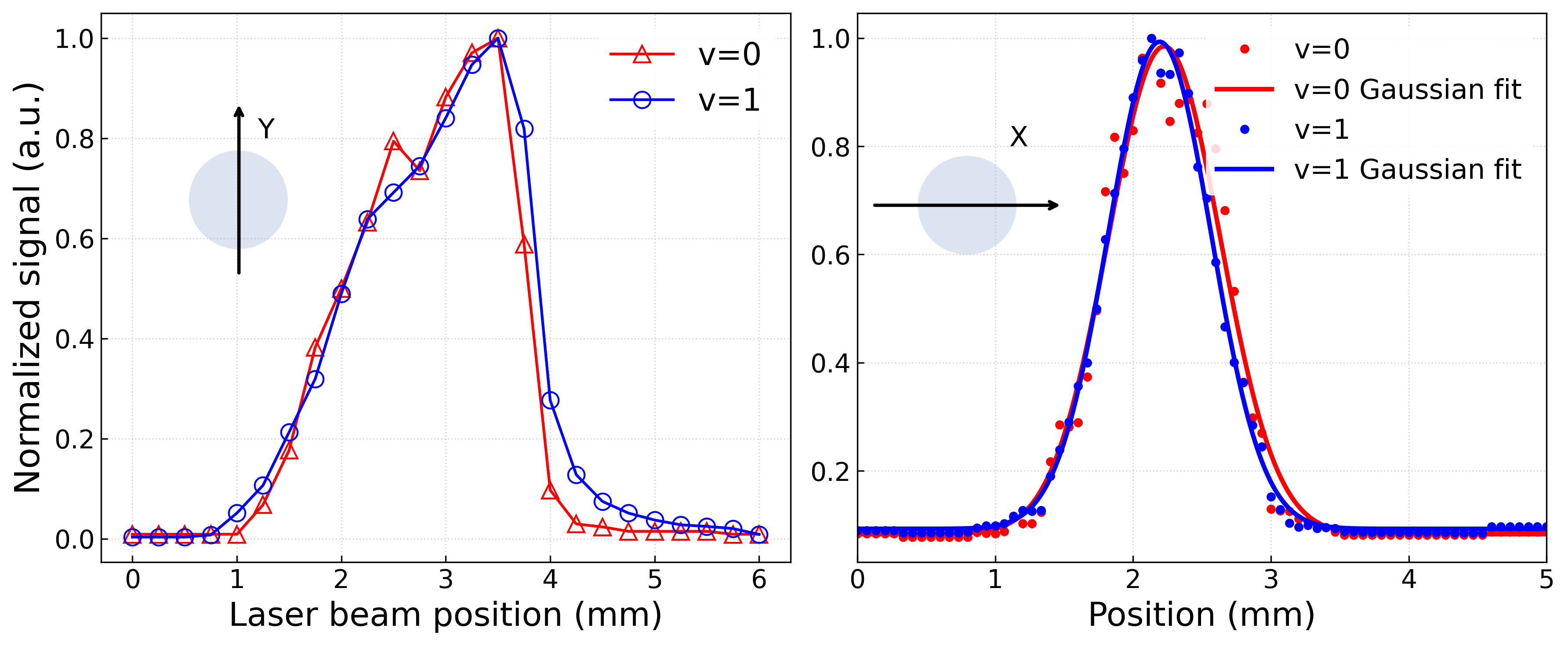}
    \end{minipage}

    \caption{Left: Schematic of QCL excitation of the molecular beam.  Center: REMPI signals of the $v$ = 0 (red curve) and $v$ = 1 (blue curve) states measured by translating the UV beam along the Y-axis. Right: Spatial overlap of the two states along the X-axis, determined from their ion images recorded sequentially under identical experimental conditions.}
    \label{fig:spacial}
\end{figure}
The spatial profiles measured along the Y and X axes for $v$ = 0 and $v$ = 1 state (Fig. \ref{fig:spacial}, left and right) show good overlap, suggesting uniform excitation across the molecular beam.

\subsection{Temporal profile}
\begin{figure}[h]
    \centering
    \includegraphics[width=1\linewidth]{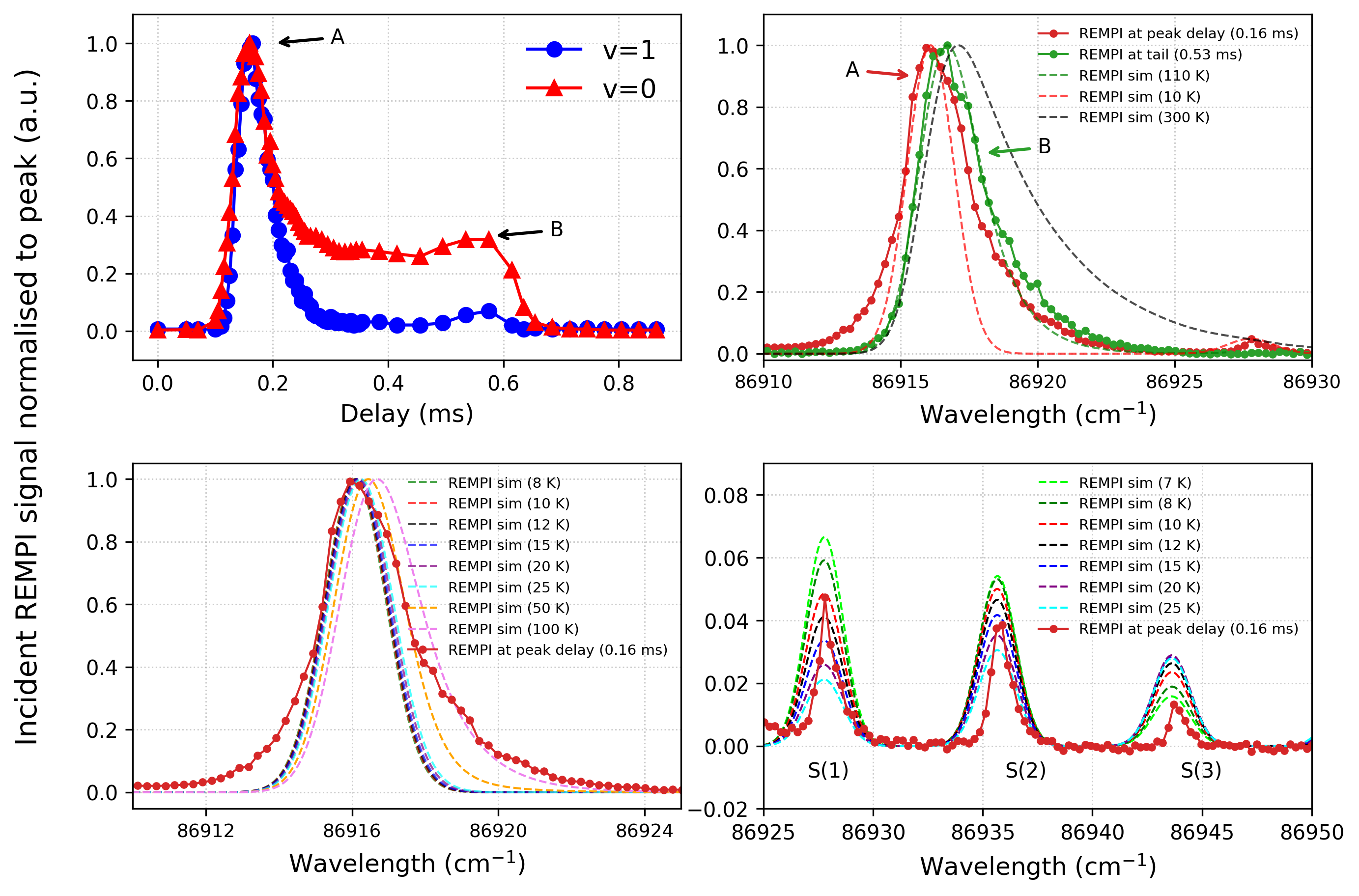}
    \caption{Top left: The red and blue curves show the temporal profiles of the $v$ = 0 and $v$ = 1 signals, respectively. Point A marks the signal peak, corresponding to the maximum beam flux, while point B marks the smaller peak at the tail region for longer delays. Top right: The red curve shows the REMPI spectrum recorded at point A, along with its simulated spectrum at 10 K (red dotted curve). The green curve shows the REMPI spectrum recorded at point B, with its simulation at 110 K (green dotted curve). The black curve represents a simulated REMPI spectrum at 300 K, for reference. Bottom left: The REMPI spectrum recorded at peak (point A) is compared with simulations at different rotational temperatures. None of the simulations fits well with the measured spectrum, which very likely due to saturation effects in the strong Q-branch. Bottom right: The red curve shows the S-branch lines extracted from the REMPI spectrum recorded at point A, along with the simulated spectra at different rotational temperatures. These lines were used to estimate the rotational temperature of the molecular beam.}
    \label{fig:temporal}
\end{figure}

The temporal profile was measured by varying the delay between the nozzle opening and the UV laser pulse (using the Q-switch) used for REMPI detection.
The measured profile (Fig. \ref{fig:temporal}, top left, red curve) peaks at 0.16 ms (point A) followed by a tail at longer delays, with a smaller peak at 0.53 ms (point B).
At the tail, the $v$ = 1 excitation efficiency is low.
The possible reasons include velocity inhomogeneity, which leads to Doppler shifts, and rotational-state inhomogeneity, which results in variations in the J = 0 population and, consequently, the overall excitation efficiency.
The measured most probable velocities at the tail and main peak are 1626 m/s and 1890 m/s, respectively. The Doppler shift arising from this velocity difference is estimated to be 0.5 MHz, which is within the laser linewidth. 
Therefore, the observed effect is primarily attributed to the differences in the rotational-state distributions.
To confirm this, REMPI spectra was measured (Fig. \ref{fig:temporal}, top right) at the peak (point A) and at the tail (point B). 
The measured spectra were fitted with REMPI simulations (PGOPHER \cite{pgopher}) revealing a significant difference in the rotation temperature.
At the tail (point B), the REMPI spectra measured agrees well with the simulation done at a rotational temperature of 110 K. 
In contrast, the simulated spectra at the peak does not fit the measured spectra for any single rotational temperature (Fig. \ref{fig:temporal}, bottom left), possibly due to signal saturation.
Therefore, the rotational temperature at the peak delay was estimated from the S-branch lines (S(1), S(2), and S(3)), as shown in the bottom right panel of Fig. \ref{fig:temporal}.
At 8 K, the simulated S(3) intensity agrees better with the experiment, while at 10–12 K, the overlap with S(1) and S(2) improves, although the S(3) intensity is overestimated. 
Above 12 K, the overall fit becomes poorer. 
Based on these comparisons, the rotational temperature at peak is estimated to be approximately in the range 8–12 K, which is significantly lower that the rotational temperature in the tail region.

\clearpage
\bibliographystyle{unsrtnat}
\bibliography{SI_ref}